\documentclass[a4paper,fleqn,usenatbib]{mnras}

\usepackage{newtxtext,newtxmath}

\usepackage[T1]{fontenc}
\usepackage{ae,aecompl}

\usepackage{graphicx}	
\usepackage{amsmath}	
\usepackage{hyperref}

\title[Nii code]{Nii: a Bayesian orbit retrieval code applied to differential astrometry} 

\author[S. Jin et al.]{
Sheng Jin$^{1,2}$\thanks{E-mail: shengjin@pmo.ac.cn}, Xiaojian Ding$^{3}$, Su Wang$^{1,2}$, Yao Dong$^{1,2}$, Jianghui Ji$^{1,2}$ 
\\
$^{1}$CAS Key Laboratory of Planetary Sciences, Purple Mountain Observatory, Chinese Academy of Sciences, Nanjing 210023, China \\
$^{2}$CAS Center for Excellence in Comparative Planetology, Hefei 230026, China\\
$^{3}$College of Information Engineering, Nanjing University of Finance and Economics, Nanjing 210007, China \\
}

\date{Accepted XXX. Received YYY; in original form ZZZ}

\pubyear{2021}

\begin{document}
\label{firstpage}
\pagerange{\pageref{firstpage}--\pageref{lastpage}}
\maketitle

\begin{abstract}
Here we present an open source Python-based Bayesian orbit retrieval code (Nii) that implements an automatic parallel tempering Markov chain Monte Carlo (APT-MCMC) strategy. Nii provides a module to simulate the observations of a space-based astrometry mission in the search for exoplanets, a signal extraction process for differential astrometric measurements using multiple reference stars, and an orbital parameter retrieval framework using APT-MCMC. We further verify the orbit retrieval ability of the code through two examples corresponding to a single-planet system and a dual-planet system. In both cases, efficient convergence on the posterior probability distribution can be achieved. Although this code specifically focuses on the orbital parameter retrieval problem of differential astrometry, Nii can also be widely used in other Bayesian analysis applications.
\end{abstract}




\section{Introduction}

Searching for exoplanets using various techniques has become increasingly important in the field of planetary science \citep{Fischer2014}.
Owing to continuous technological advancements, past and ongoing surveys have detected $\sim$ 6,000 confirmed exoplanets and candidates\citep{Cumming2008,Borucki2011,Mayor2011,Cassan2012,Marcy2014,Thompson2018}.
This large sample provides the basis for thoroughly understanding the processes by which planets form and evolve \citep{Ida2004,Mordasini2012,Bitsch2015,Liu2020,Zhang2020}.

The astrometry method employed to detect exoplanets is implemented by precisely measuring the tiny wobbles of a star caused by the gravitational pull exerted by one or more surrounding planets.
The first formal astrometric calculation for an extrasolar planet was made in 1855 for the star 70 Ophiuchi \citep{Jacob1855},
but unfortunately subsequent observations discovered that the signal was entirely due to systematic errors in the visual measurements in the 19th century \citep{Heintz1988}.
The precision required to detect a planet using the astrometry technique is extremely high; thus far only a handful of previously discovered exoplanets 
have been successfully  recharacterized by astrometric method \citep{Benedict2002,Snellen2018,Feng2019}.
In addition, the astrometric method is sensitive to exoplanets with large orbits; thus long observation times are necessary to complete the orbits of planets with long orbital periods.
The GAIA mission is expected to reach a magnitude-dependent accuracy of $\sim$ 10 $\mu$as and is expected to find thousands of Jovian exoplanets up to 500 parsecs from the Sun via astrometry \citep{Perryman2014}.

To discover terrestrial planets, the accuracy of the astrometry technique must reach $\sim$ 1 $\mu$as.
A feasible way to achieve such a high accuracy is to develop a space-based differential astrometry telescope.
The principle of differential astrometry is that it measures and compares the relative offset angles between a target and several distant reference stars, not the absolute position of the target in the celestial sphere.
Using narrow-angle astrometric observations, such space-based instruments can differentially detect the reflex motion of the target star due to the presence of its planets with very high accuracy \citep{Catanzarite2006,Unwin2008,Goullioud2008}.
Historically, a series of space missions have been proposed to search for exoplanets using differential astrometry; examples include the Space Interferometry Mission PlanetQuest (SIM PlanetQuest) \citep{Catanzarite2006,Unwin2008,Tanner2010}, the Nearby Earth Astrometric Telescope (NEAT) mission \citep{Malbet2012} and the Search for Terrestrial Exo-Planets (STEP) mission \citep{Chen2013,Liu2018}.
Several follow-up successions of these projects are in progress, namely, the Theia Space Observatory \citep{Malbet2016,Theia2017}, the Habitable ExoPlanet Survey (HEPS) mission \citep{Yu2019} and the Closeby Habitable Exoplanet Survey (CHES) mission \citep{Ji2020}.

Space-based missions using the differential astrometry technique can provide the offset angles between a target star and its reference stars in a time series.
Thus, a successful orbit retrieval model needs to accurately fit the planetary mass and orbital parameters with these time series data.
Such a process is similar to the fitting of planetary orbital parameters in the radial velocity method and the absolute astrometry method; Bayesian methods, especially Markov chain Monte Carlo (MCMC) methods, are widely employed for these tasks \citep{Gregory2005,Gregory2005b,Ford2006,Balan2009,Gregory2011,Schulze2012,Eastman2013,Diaz2014,Feroz2014,Ranalli2018,Brandt2021}.
In the case of differential astrometry, additional attention should be paid to the error sources related to the special observation strategy and the use of multiple reference stars \citep{Liu2018}.

Here, we provide a Bayesian orbit retrieval framework for potential differential astrometry missions in an open source code known as Nii\footnote{\url{https://github.com/shengjin/Nii.git}}.
To handle the high-dimensional model of a planetary system, we implement the strategy of automatic parallel tempering Markov chain Monte Carlo) (APT-MCMC) strategy \citep{Liu2001,Gregory2005,Gregory2005b}.
Our code implements two special treatments.
First, in the analysis of the observation errors, we consider the effect of using multiple-reference stars in the differential astrometry method.
Second, in the control system utilized to automate the selection of the sampling step sizes for all the parallel Markov chains, we adopt a different scheme from the existing models \citep{Gregory2005,Gregory2005b}. 
Ultimately, we find that the Nii code can guarantee efficient convergence on the high-dimensional posterior distributions encountered in orbit retrieval problems.

The remainder of this paper is organized as follows. 
Section \ref{sec:simu} describes the forward model that is used to simulate the observations acquired during a differential astrometry mission.
Section \ref{sec:fit} elaborates on the APT-MCMC model built to retrieve orbital parameters.
Section \ref{sec:case} presents two applications of the Nii code in the retrieval of orbital parameters from a single-planet system and a dual-planet system.
Section \ref{sec:sum} presents a brief summary. 

\section{Forward model: Simulated observations}

\label{sec:simu}

The relative movements of  nearby bright stars calibrated using several reference stars can be measured with sub-microarcsecond precision using a space-based differential astrometry telescope equipped  with a Michelson interferometer \citep{Unwin2008,Shao2009,Malbet2012,Theia2017,Ji2020}.
The technique utilized for such missions is known as micropixel image position sensing, which measures the differential motion between the centroids of a target star and a reference star \citep{Nemati2011,Zhai2011}.
In this technique, the detector pixel response functions in Fourier space are characterized using laser metrology, and these response functions are further used to construct a point spread function (PSF).    
The derived PSF is resampled at different relative locations from the original position of the centroid of the target star ($\Delta x_{c}$s, $\Delta y_{c}$s) and compared with the PSF of a second image taken at a later time.
Then, the resampled PSF at the location ($\Delta x_{c}$, $\Delta y_{c}$) that best matches the PSF of the second image can help determine the displacement of the centroid of the target star based on a reference star \citep{Nemati2011,Zhai2011}.
With this method, an astrometric accuracy of $6\times 10^{-5}$ pixels can be achieved \citep{Crouzier2016}.
The space missions proposed for the survey of exoplanets based on differential astrometry have evolved in recent years\citep{Theia2017,Yu2019,Ji2020}.

In a space-based differential astrometry mission, the observation output of each target star in the operating cycle is a time series of the relative movements between the target star and several selected distant reference stars.
These relative movements contain multiple parts.
First, the relative displacements caused by the proper motions and  parallaxes of the target star and the reference stars are the main contributors to these movements.
In the Nii code, the proper motions of the stars are set as necessary input parameters in the simulated observations, and the parallaxes of the stars are can also be included in the same module.
Consequently, we can subtract the relative displacements caused by the proper motions and parallaxes of the target and reference stars based on these input parameters. 
However, this assumes that we accurately know these parameters, which is unlikely in reality.
When dealing with real observations, we must conduct a special analysis of the proper motions and parallaxes of the stars within a particular system.
We should also investigate complex situations, for example,  a potential signal with a period close to a year, as such a signal would produce an astrometric signal that could constructively or destructively interfere with the measured parallax and severely reduce the precision of the simulated measurements.
For simplicity,  as this work focuses on the retrieval of planetary orbital parameters, we assume that the proper motions and parallaxes of the stars are accurately known and can be directly subtracted from the detected relative movements.

If the target star hosts a planet, then the remaining periodic signals after the proper motions and parallaxes of all the stars are subtracted may correspond to either the planetary system of the target star or the wobbles of a reference star. 
Such periodic signals created by planetary systems present as movements in rectangular coordinates on the focal plane at different observation time points.
As a source of uncertainty, these periodic signals may be related to  either the planetary system of the target star or  the planetary system of a reference star.
Because the selected reference stars are far from the target star, only the wobbles generated by massive Jovian planets orbiting these reference stars can be detected. 
Therefore, the embodiment of such uncertainty is that we cannot distinguish the source of the detected periodic signals between a terrestrial planet orbiting the target star or a Jovian planet orbiting one of the reference stars.
The solution of this problem is the observation scheme of using multiple-reference stars in a space-based differential astrometry mission.
A disturbance caused by the planetary system of one reference star cannot appear in the differential astrometric signal obtained using another reference star.
Thus, the source of a periodic signal can be identified by comparing the fitting results derived by using different reference stars.
Similarly, such a cross-comparison approach can reduce the contamination of the fitting results arising from the errors in the proper motion or parallax of a particular reference star.

In a simple case in which the proper motions and parallaxes of the target and reference stars are accurately removed, the output of an observation pipeline for a target star with a planetary system can be reduced to periodic changes in right ascension and declination around the centre of mass of the system.
Considering a target star with a single-planet system, following the approach in the radial velocity detection method \citep{Gregory2005}, the changes in right ascension and declination over time can be expressed as
\begin{equation}
	\label{eq:deltapie}
$$
\begin{cases}
	\Delta \eta'(t)= \Delta \eta(t) + \epsilon_{\eta}(t) + \epsilon_{x} \\
	\Delta \delta'(t)= \Delta \delta(t) + \epsilon_{\delta}(t) + \epsilon_{x} \\
\end{cases}
$$
\end{equation}
where $\Delta \eta'(t)$ and $\Delta \delta'(t)$ are the measured changes between each pair of the target star and a reference star in right ascension and declination at the instant of time $t$,
$\Delta \eta(t)$ and $\Delta \delta(t)$ are the changes in right ascension and declination due to the presence of a planet, 
$\epsilon_{\eta}(t)$ and $\epsilon_{\delta}(t)$ are the noise components due to known but unequal measurement errors with an assumed Gaussian distribution,
and $\epsilon_{x}$ represents any additional unknown measurement errors plus any real signal that cannot be described by $\Delta \eta$ and $\Delta \delta$ (e.g., a signal caused by another planet or the unexpected movement of any reference star). 

Suppose the planet orbiting the target star has a mass $M_{\mbox{p}}$ and orbital elements 
$(a, e, i, \omega, \Omega, M_0)$, where $a$ is the orbital semimajor axis, $e$ is the eccentricity, $i$ is the inclination, 
$\omega$ is the argument of periapsis, $\Omega$ the longitude of the ascending node, and $M_0$ is the mean anomaly at a reference time.
Note that such an orbital parameter set will lead to ambiguity in the fitted $\omega$ and $\Omega$, especially for orbits with small eccentricity and small inclination.
One solution is to use equinoctial orbit elements \citep{Brouche1972}.
For simplicity, the orbit retrieval in Nii employs Keplerian elements since the $M_{\mbox{p}}$, $a$, $e$, $i$, and the phase curve of the orbit are the only information  expected to be  known. 
Nevertheless, although $\omega$ is needed to predict the radial velocity curve and $\Omega$ defines the position angles of the orbits of both the planet and the host star around the photocentre of the sky,
these parameters are not relevant when determining the mass of the planet and orbital period, the two main objectives of differential astrometry missions \citep{Ji2020}.

For precise planetary mass determination, the stellar mass should be an additional free parameter so the uncertainty could be marginalized over through Bayesian sampling.
Given that the semimajor axis of an astrometric orbit is inversely proportional to the stellar mass, the effect of using a fixed star mass should be small.
Thus in this work, we assume the stellar mass is precisely known. 
We will parameterize the stellar mass in further development of the Nii code.

To simulate the differential astrometric signal obtained from an observation pipeline, the elliptical orbital motion of the target star induced by the planet should be projected onto the observation plane.
The Thiele--Innes elements, $(A,B,F,G)$, are used to calculate the wobble in rectangular coordinates \citep{Thiele1883,Alzner2004} as follows:
\begin{equation}
	\label{eq:thiele}
$$
\begin{cases}
 A=\alpha(\cos \Omega \cos \omega - \sin \Omega \sin \omega \cos i) \\
 B=\alpha(\sin \Omega \cos \omega + \cos \Omega \sin \omega \cos i) \\
 F=\alpha(-\cos \Omega \sin \omega - \sin \Omega \cos \omega \cos i) \\
 G=\alpha(-\sin \Omega \sin \omega + \cos \Omega \cos \omega \cos i)
\end{cases}
$$
\end{equation}
where $\alpha$ is the maximum astrometric amplitude of the target star (with a mass of $M_{\ast}$ at a distance $d$ from the solar system) due to the reflex motion in the presence of the planet given by
\begin{equation}
\label{eq:alpha}
	\alpha = 3 \left(\frac{M_{\mbox{p}}}{1M_{\oplus}}\right)  \left(\frac{a}{1\mbox{AU}}\right)
	\left(\frac{M_{\ast}}{1M_{\odot}}\right)^{-1}
	\left(\frac{d}{1\mbox{pc}}\right)^{-1}  \mu\mbox{as}.
\end{equation}
For an Earth-like planet in the habitable zone of a solar-like star at 10 parsecs, a typical value of $\alpha$ is 0.3 $\mu\mbox{as}$.

Given these Thiele--Innes elements, the reflex motion of the target star in right ascension and declination due to the presence of a single planet can be solved by
\begin{equation}
\label{eq:delta}
$$
\begin{cases}
	\Delta \eta(t)= A X(t) + F Y(t) \\
	\Delta \delta(t) = B X(t) + G Y(t) 
\end{cases}
$$
\end{equation}
where
\begin{equation}
	\label{eq:XYE}
$$
\begin{cases}
	X(t) = \cos E(t)-e       \\ 
	Y(t) = \sqrt{1-e^{2}} \sin E(t) \\
	E(t) - e \sin E(t) =  \frac{2\pi}{P}(t-T) 
\end{cases}
$$
\end{equation}
where $E$ is the eccentric anomaly that can be determined from the mean anomaly $M_0$
and $T$ the time of passage through the periapsis.

For a specific system and a specific time series $t$, we first simulate the $\Delta \eta(t)$ and $\Delta \delta(t)$ of the target stars using Equations \ref{eq:thiele}, \ref{eq:alpha}, \ref{eq:delta} and \ref{eq:XYE} with a planet characterized by the parameter set $(a, e, i, \omega,\Omega, M_0, M_{\mbox{p}})$.
Then, we add the noise components $\epsilon_{\eta}(t_i)$ and $\epsilon_{\delta}(t_i)$ for each instant of time $t_i$ in the time series, and we assume that all $\epsilon_{\eta}(t_i)$ and $\epsilon_{\delta}(t_i)$ follow a Gaussian distribution with a mean of 0 and standard deviations of $\sigma_{\eta}(t_i)$ and $\sigma_{\delta}(t_i)$ respectively, which depend on the accuracy of the differential astrometric measurement.

There are two scenarios that may lead to additional measurement errors of $\epsilon_{x}$. 
In the first scenario, there is another unknown planet in the simulated system.
In the second,  one of the reference stars wobbles due to a planet or a companion star, 
which leads to unexpected errors in the differential astrometric signal.
The difference between these two scenarios is that the additional errors caused by the wobbles of a specific reference star do not appear in the differential astrometric measurement of another reference star.

In this work we simulate a host star with a mass of 1 $M_{\odot}$ that is located 3 parsecs from the Earth.
We calculate the planetary system of this star in the following two scenarios:
\begin{enumerate}
\item  The target star is within a single-planet system.
\item  The target star is within a dual-planet system.
\end{enumerate}

The simulated space mission uses CHES as a prototype \citep{Ji2020}.
The mission time is 5 years, during which approximately 300 measurements are acquired for each target star.
The field of view of the CHES mission is 0.44$^{\circ} \times$ 0.44$^{\circ}$, 
in which the relative offset angles between the host star and at least 8 reference stars 
are measured using the differential astrometry technique.
CHES aims to achieve an accuracy of $\sim$ 1 $\mu\mbox{as}$.
This information is summarized in Table \ref{tab:fit0}.

We generate the simulated observation signals for these two scenarios using Equation \ref{eq:deltapie} and adopt the method described in Section \ref{sec:fit} to fit the orbital parameters of the planetary system of the target star.

\begin{table}
\caption[Parameters of all the models]
{Parameters of the host star and the simulated observation.}
\begin{tabular}{lr}
\hline
\hline
Stellar mass ($M_{\odot}$)  & 1.0 \\
Distance from the Earth (parsec) & 3.0 \\
Field of view &  0.44$^{\circ} \times$ 0.44$^{\circ}$ \\
Number of reference stars &  8 \\
Mission duration  (year) & 5 \\
Number of observations & 300 \\
Simulated Gaussian measurement error ($\mu\mbox{as}$) & 1.0 \\
\hline
\end{tabular}
\label{tab:fit0}
\end{table}
\normalsize

\section{Bayesian orbit retrieval}
\label{sec:fit}

\subsection{Bayesian inference}

The orbit retrieval procedure in Nii for a planet revolving around a target star is based on Bayesian analysis \citep{Bayes1763}.
Let $\theta$ be the unknown parameter set that we want to estimate with a specific model $M$.
Then, the posterior probability distribution for $\theta$ after we obtain  observation data $D$ is of the following form:
\begin{equation}
p(\theta|D,M)=\frac{p(D|\theta,M)\,p(\theta|M)}{p(D|M)}\,
\label{eqn:bayes1}
\end{equation}
where $p(D|\theta,M)$ is the likelihood reflecting the probability of generating the particular observation data $D$ if the parameters in the model $M$ are equal to $\theta$ and 
$p(\theta|M)$ is the prior distribution of $\theta$ representing our beliefs across different values of the parameters before observing the data.
The denominator ${p(D|M)}$ is the normalization constant ensuring that the posterior distribution integrates to one.
Given observation data $D$, we can simply omit the calculation of the denominator and evaluate the posterior distribution as
\begin{equation}
p(\theta|D,M) \propto p(D|\theta,M)\,p(\theta|M)\,
\label{eqn:bayes2}
\end{equation}

The main goal of orbit retrieval is to estimate the posterior probability distribution in Equation \ref{eqn:bayes2} by sampling.
The point estimate of each parameter is set as the posterior mean because it is representative of the central position of the posterior distribution and mathematically accounts for the measure from a measure-theoretic perspective. 
Nii implements the APT-MCMC method \citep{Liu2001,Gregory2005}, which is introduced in detail in Section \ref{sec:ptmcmc}. 
The underlying sampling algorithm of Nii's MCMC strategy is Metropolis--Hastings \citep{Metropolis1953,Hastings1970}.

\subsection{Prior distributions}

\begin{table}
\caption[Choice of Priors]
{Prior distributions of all the model parameters.}
\begin{tabular}{ccccc}
\hline
\hline
Parameter & Prior  & Mathematical Form & Min & Max \\ 
\hline
$P ({\mathrm {days}})$& Jeffreys & $\frac{1}{P\,\ln\big(\frac{P_{\mathrm {max}}}{P_{\mathrm {min}}}\big)}$ & 0.5 & 3650 \\ 
&  &  &  &  \\ 
$M_{\mbox{p}} (M_{\oplus})$ & Jeffreys & $\frac{1}{M_{\mathrm p}\,\ln\big(\frac{M_{\mathrm {max}}}{M_{\mathrm {min}}}\big)}$ & 0.1 &3000 \\ 
&  &  &  &  \\ 
$e$ & Uniform & 1 & 0 & 1 \\ 
&  &  &  &  \\ 
$cosi$ & Uniform & 0.5 & 1 & -1 \\ 
&  &  &  &  \\ 
$\Omega$ & Uniform & $\frac{1}{2\pi}$ & 0 & $2\pi$ \\ 
&  &  &  &  \\ 
$\omega$ & Uniform & $\frac{1}{2\pi}$ & 0 & $2\pi$ \\ 
&  &  &  &  \\ 
$M_{\mathrm 0}$ & Uniform & $\frac{1}{2\pi}$ & 0 & $2\pi$ \\ 
&  &  &  &  \\ 
$\epsilon_{x} (\mu\mbox{as})$ & Mod. Jeffreys & $\frac{(\epsilon_{x}+\epsilon_{xa})^{-1}}{\ln\big(\frac{\epsilon_{xa}+\epsilon_{x_{\mathrm {max}}}}{\epsilon_{xa}}\big)}$ & 0 ($\epsilon_{xa}=1$) & $100$\\
\hline
\end{tabular}
\label{tab:prior}
\end{table}
\normalsize

The choice of priors is very important in Bayesian inference since improper priors can produce misleading results. 
We assume that the prior distribution of each parameter is independent of that of each other parameter and set the prior distribution in a similar way to the radial velocity detection method \citep{Gregory2005}.

In the case that the mass of the host star is known, every additional planet needs 7 more parameters to determine its orbit.
In addition to the parameters describing planetary orbits, there is also an $\epsilon_x$ that represents any additional unknown errors.
Table~\ref{tab:prior} shows the choice of priors for each parameter and their boundaries. 
We choose uniform priors for $e$, $\omega$, $\Omega$, and $M_0$, 
and the inclination $i$ is set to be uniform in $cosi$. 
For $M_{\mathrm p}$, the planetary orbital period $P$ and $\epsilon_x$, a uniform prior would be inappropriate since it would strongly favour a parameter range with larger values.
We choose the Jeffreys prior for $M_{\mathrm p}$ and $P$ and a modified Jeffreys prior for $\epsilon_x$ \citep{Gregory2005}. 

For a simple model where the target star hosts only one planet, the joint prior distribution for the model parameters, assuming independence, can be written as
\begin{equation}
\begin{split}
    p(\theta|M)\, = \,\,  &p(P|M)\,p(M_{\mathrm p}|M)\,p(e|M)\,p(cosi|M)\,p(\Omega|M)\\
 & \quad \quad \times p(\omega|M)\,p(M_0|M)\,p(\epsilon_0|M)
\label{eqn:priors}
\end{split}
\end{equation}

\subsection{Likelihood function}

One of the features of the Nii code is that it considers the effect of using multiple-reference stars in the analysis of the observation errors. 
At all moments in an observation time series, we use the relative position angles between a target star and all the reference stars to calculate the standard deviation of the Gaussian observation error at each instant of time.
Then, we separately pair the target star and each reference star to perform the orbit retrieval process, in which all the measurements of the wobbles between the target star and each reference star in the observation time series are used.
Depending on the fitting results obtained using different reference stars, an additional voting process on the final fitted parameters may be required.
Thus, the effect of using multiple reference stars in the differential astrometry method is to restrict the standard deviation at a single measurement time point and to cross-validate the final fitting results.

Suppose there are a total of $N$ measurements in an observation period $t$, and each measurement yields $\epsilon_{\eta}(t_i)$ and $\epsilon_{\delta}(t_i)$ between the target star and every reference star.
If there are $N_{\mathrm {ref}}$ reference stars in total, then both $\epsilon_{\eta}(t_i)$ and $\epsilon_{\delta}(t_i)$ have $N_{\mathrm {ref}}$ elements.
Assuming at each instant of time $t_i$ the measurement errors for the $N_{\mathrm {ref}}$ reference stars follow a Gaussian distribution with a mean of 0, we can calculate the standard deviation $\sigma_{\eta}(t_i)$ and $\sigma_{\delta}(t_i)$ in the Gaussian distribution based on the $N_{\mathrm {ref}}$ elements for both $\epsilon_{\eta}(t_i)$ and $\epsilon_{\delta}(t_i)$.

Since in differential astrometry a measurement gives the offset angles in two directions, namely, right ascension and declination, there are a total of $2N$ measurements between the target star and each reference star in the observation period $t$.
Nevertheless, this is only an ideal situation, and some data may be missing in real cases.
The likelihood function for the $2N$ measurements of the offset angles between the target star and each reference star is given by \citep{Gregory2005,Bishop2006}
\begin{equation}
\begin{split}
	p(D|\theta,M)= \, A\, &\exp \Biggl\{-\sum_{i=1}^{N}\frac{\big[ \Delta \eta'(t_i) -\Delta \eta(t_i) \big]^2}{ 2\big[\sigma_{\eta}(t_i)^2 + \epsilon_{x}^2\big] }\Biggr\}\, \\
	& \times \exp \Biggl\{-\sum_{i=1}^{N}\frac{\big[ \Delta \delta'(t_i) -\Delta \delta(t_i) \big]^2}{ 2\big[\sigma_{\delta}(t_i)^2 + \epsilon_{x}^2\big] }\Biggr\}\,
\label{eqn:lik}
\end{split}
\end{equation}
where
\begin{equation}
	A=(2\pi)^{-N}\prod_{i=1}^{N}\Big[\sigma_{\eta}(t_i)^2+\epsilon_{x}^2\Big]^{-1/2}\,\prod_{i=1}^{N}\Big[\sigma_{\delta}(t_i)^2+\epsilon_{x}^2\Big]^{-1/2}\,
\label{eqn:likA}
\end{equation}

Equation \ref{eqn:lik} demonstrates that the $N_{\mathrm {ref}}$ reference stars are used together only to determine the standard deviation of the distribution of the measurement errors at each instant of time $t_i$.
In the orbit retrieval process, we separately fit the orbital parameters using the offset angles between the target star and each reference star in the entire time series $t$.
This approach can reduce a sizable number of parameters in $\theta$ because each reference star has a different unknown error represented by $\epsilon_x$.
Finally, we summarize the fitting results obtained from different reference stars to evaluate the reliability of the inferred orbital parameters.

\subsection{Implementation of PT-MCMC}
\label{sec:ptmcmc}

The posterior distribution given by Equation \ref{eqn:bayes2} is highly multimodal, and an MCMC process can become stuck in many local optimal solutions and fail to fully explore the global distribution.
One solution to this problem is referred to as simulated tempering \citep{Geyer1995,Gregory2005b}, in which flatter versions of the target posterior distributions are generated using a temperature parameter $\beta$:  
\begin{equation}
	\widehat{p}(\theta|D,\beta,M) \propto p(D|\theta,M)\,p(\theta|M)^{\beta}\,, \quad {\mathrm {for}} \,\, \, 0<\beta<1
\label{eqn:temp}
\end{equation}
where $\beta$ varies from 1, corresponding to the coldest original posterior distribution, to 0, corresponding to the hottest distribution that is completely flat.
Using a flat posterior distribution described by Equation \ref{eqn:temp}, an MCMC sampler with a suitable $\beta$ can easily escape from local optimal solutions.

An attractive technique, parallel tempering (PT), runs several Markov chains with different $\beta$ values in parallel \citep{Liu2001,Gregory2005b}.
In a PT-MCMC run, one chain with $\beta = 1$ corresponds to the original posterior distribution, and other chains with different smaller $\beta$ values form a set of ladders leading to different higher temperature distributions.
At random intervals, e.g., a state $s$, a pair of adjacent chains with temperature parameters $\beta_i$ and $\beta_{i+1}$ are chosen at random, and their current parameter states, $\theta_{s,i}$ and $\theta_{s,i+1}$, are interchanged with a probability of
\begin{equation}
	r = {\mathrm {min}} \Biggl[\,1, \, \frac{ \widehat{p}\big(\theta_{s,i+1}|D,\beta_i,M\big) \, \widehat{p}\big(\theta_{s,i}|D,\beta_{i+1},M\big)}{ \widehat{p}\big(\theta_{s,i}|D,\beta_i,M\big) \, \widehat{p}\big(\theta_{s,i+1}|D,\beta_{i+1},M\big)} \, \Biggr]\,
\label{eqn:pltemp}
\end{equation}
Such swaps of parameter states allow parallel chains to exchange their sampling areas frequently, facilitating the convergence of global sampling.

In our orbit retrieval model, 8 MCMC chains are run in parallel with a set of $\beta = \{0.01, 0.02, 0.05, 0.1, 0.25, 0.5, 0.75, 1.0\}$.
A proposed swap is randomly evaluated using Equation \ref{eqn:pltemp} every 10 to 30 iterations.

\subsection{Automatic Gaussian proposal distributions: APT-MCMC}

One key factor that determines the completeness of MCMC sampling is the Metropolis--Hastings step size of each parameter $\theta_i$ (characterized by the standard deviation of the Gaussian proposal distribution $\sigma_{\theta_i}$).  
If the step sizes are too small, it is difficult for the MCMC sampler to cover the entire parameter space and find the optimal solution since the obtained posterior density is highly dependent on the starting location of the chain. 
In contrast, if the step sizes are too large, the MCMC sampler rejects the majority of proposals, and only a few effective samples are obtained.
For a large number of parameters, an ideal combination of step sizes leads to an acceptance rate of $\sim$ 23.4\% \citep{Gelman1997}.

In a high-dimensional orbit fitting model, it is difficult to find an ideal combination of sampling step sizes for all PT Markov chains.
First, the ideal step sizes for different parameters $\theta_i$ are unique and change under different $\beta$ values.
Second, when exploring different regions of the target posterior distribution, the ideal set of step sizes that is compatible with the shape of the local posterior distribution should also be adjusted accordingly.
Therefore, in the case of the simplest single-planet model with 8 parameters, a PT-MCMC program with 8 chains with different $\beta$ values needs to tune 64 different step sizes in real time to reach the optimal acceptance rate, which is almost impossible to perform manually.

The Nii code adopts a different automatic control system from the existing model \citep{Gregory2005,Gregory2005b}. 
In our APT-MCMC scheme, the step sizes of all the parameters of all the PT Markov chains are adjusted dynamically in real time after an initial burn-in stage.  
At the initial burn-in stage, we set the standard deviation of the Gaussian proposal of each parameter, $\sigma_{\theta_i}$ , to $10\%$ of the parameter's prior range to ensure that the sampler can access all the high-density regions of the posterior distribution.
However, these initial step sizes result in extremely low acceptance rates for Markov chains.
Thus, after the burn-in stage, the Nii code monitors the acceptance rate of each Markov chain in real time and dynamically adjusts the standard deviations of the Gaussian proposals of all the parameters for the chains with a low acceptance rate.
This process is implemented by dividing the entire Markov chains into many short cycles and adding many tuning stages for the chains with poor acceptance rates after each short cycle.
In these tuning stages, the program separately adjusts the $\sigma_{\theta_i}$ of each parameter with a scaling array and observes the corresponding changes in the acceptance rate.
Then, a new combination of step sizes is determined by combining two factors: the sensitivity of the acceptance rate to each $\sigma_{\theta_i}$ and the deviations between the ideal acceptance rate and the acceptance rates obtained by all the scaled parameters in the tuning stage.
The strategy here is to select from the tuned parameter matrix the parameter set that can result in an acceptance rate that is closest to the ideal rate of $23.4\%$ in the local posterior distribution being sampled.
Although this method cannot obtain the ideal acceptance rate of $\sim 23.4\%$, it is sufficient for efficient convergence. 
In our test runs, an APT-MCMC sampler can converge on the posterior distribution within 500,000 iterations.

\subsection{Gelman-Rubin convergence diagnostics}

We use the Gelman-Rubin criterion, $R_c$, to assess whether the APT-MCMC sampling process has converged \citep{Gelman1992}.
Gelman-Rubin diagnostics use an analysis of both the between-chain variance and within-chain variance of several independent Markov chains to assess convergence. 
Suppose there are $N$ Markov chains of length $L$. 
Let $\theta_{ij}$ denote the $i$th of the $L$ iterations of $\theta$ in the $j$th simulated chain, $\bar{\theta}_j$ denote the sample posterior mean of chain $j$, and $\bar{\theta}$ be the overall sample posterior mean of the $N$ chains.
The between-chains variance, $B$, is given by
\begin{equation}
    B = \frac{L}{N-1} \sum_{j=1}^{N}(\bar{\theta}_j - \bar{\theta})^2
\label{eqn:btchvar}
\end{equation}
The within-chain variance, $W$, is given by
\begin{equation}
    W = \frac{1}{N} \sum_{j=1}^{N} [\frac{1}{L-1} \sum_{i=1}^{L}(\theta_{ij}-\bar{\theta}_j)^2] 
\label{eqn:wichvar}
\end{equation}
Then, we can obtain an unbiased estimator of the marginal posterior variance of $\theta$, $var(\theta)$, by taking the weighted average of $B$ and $W$: 
\begin{equation}
    var(\theta) = (1-\frac{1}{L})W + \frac{1}{L}B
\label{eqn:vartheta}
\end{equation}
Since the initial value of $\theta$ is set randomly during MCMC sampling, $var(\theta)$ should overestimate the true marginal posterior variance. 
On the other hand, $W$ tends to underestimate the within-chain variance in the early phase of MCMC sampling.
As the sampling process gradually converges, both $var(\theta)$ and $W$ stabilize at the true variance of $\theta$.
Thus, the Gelman-Rubin criterion uses the potential scale reduction factor, $R_c$, as a convergence diagnostic
\begin{equation}
    R_c = \sqrt{\frac{var(\theta)}{W}}
\label{eqn:gelrub}
\end{equation}
In practice, the values of the Gelman-Rubin criterion should be less than 1.2 for model parameters to declare convergence in MCMC sampling, in more rigorous diagnostics, the values of $R_c$ are required to be less than 1.1 \citep{Gelman1992,Brooks1998}.

To assess the convergence of the model parameters in our Bayesian orbit retrieval framework, we run a batch of independent APT-MCMC sampling tests to calculate the Gelman-Rubin criterion.
For simplicity, we carry out this process for only one reference star.
Because all the reference stars are modelled as stable systems with no observable strong perturbations, the simulated differential astrometric signals derived from different reference stars all give the relative displacement of the target star. 
Only the random error is different.
In this case, the $R_c$ value obtained from multiple independent APT-MCMC sampling runs using the simulated observation of one reference star can help us evaluate the convergence of the sampling process using the data of other reference stars.

\section{Applications}
\label{sec:case}

\begin{figure*}
	\includegraphics[width=7.5in]{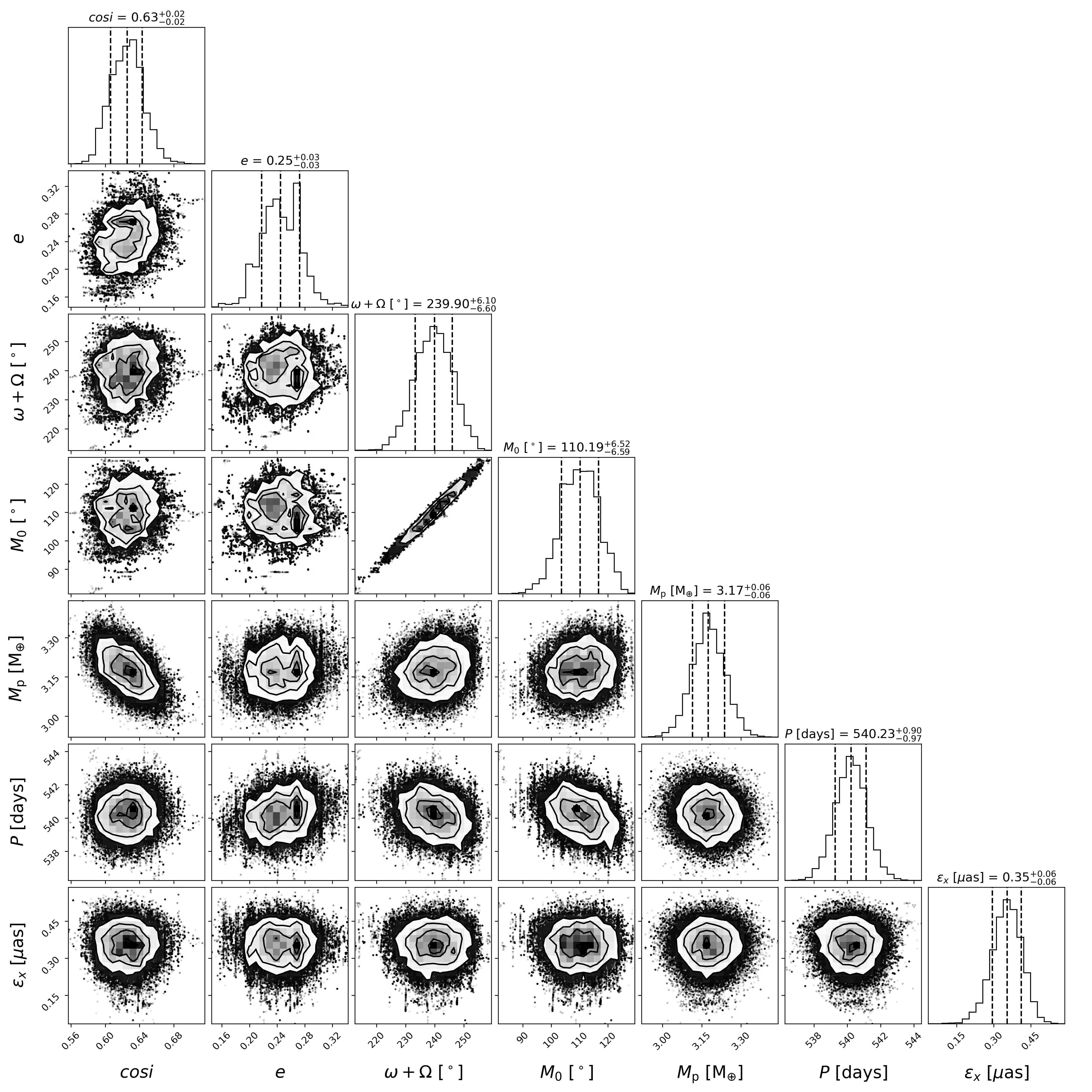}
    \caption{
        Corner plot of the posterior distributions of $a$, $e$, $cosi$, $\omega$+$\Omega$, $M_0$, $M_{\mbox{p}}$, and $P$ obtained using the differential astrometric signals of one reference star in the single-planet model.
        The corresponding marginal posterior means and one standard deviation values (and these values obtained using the other 7 reference stars) are given in Table \ref{tab:fit1p}. 
    }
    \label{fig:1b}
\end{figure*}

\begin{figure*}
	\includegraphics[width=6.5in]{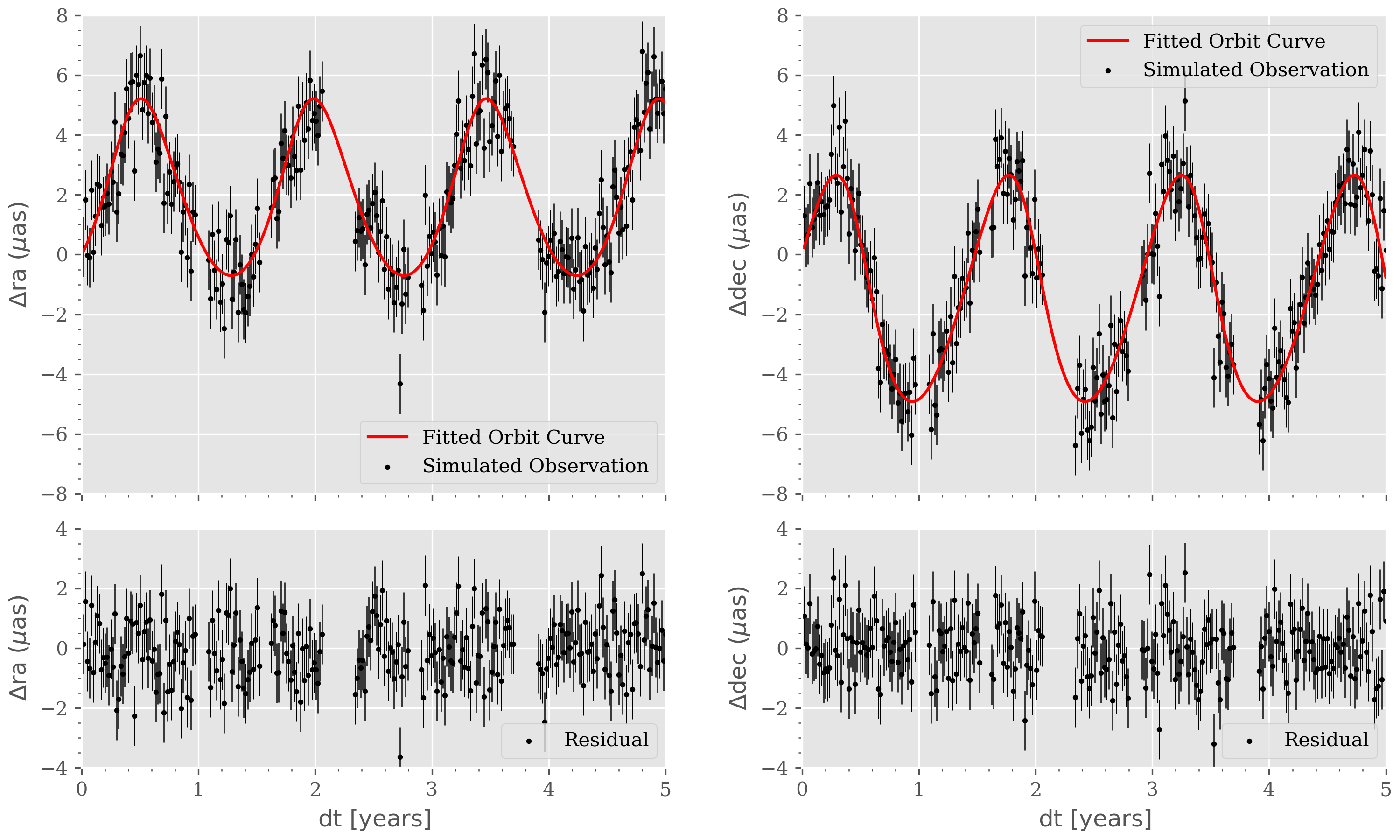}
    \caption{
        Comparison between the simulated observations of the single-planet model and the theoretical signals generated by the fitted parameters corresponding to Figure \ref{fig:1b}.
	    The points with error bars show the simulated differential astrometric observations, and the red lines signify the theoretical signals corresponding to our fitted parameters.
            The bottom panels show the residuals. 
    }
    \label{fig:1a}
\end{figure*}

\begin{table*}
\caption[Gelman-Rubin criterion of each parameter]
    {Gelman-Rubin criterion of each parameter calculated using 8 independent APT-MCMC samplings based on the simulated observations of one reference star.} 
\begin{tabular}{cccccccccc}
\hline
\hline
    Parameter & run1 & run2 & run3 &run4 &run5 &run6 &run7 &run8 & $Rc$ value \\
    $P ({\mathrm {days}})$ & 540.22$\pm0.94$ & 540.35$\pm0.92$ & 540.18$\pm0.99$ & 540.11$\pm0.94$ & 540.23$\pm0.97$ & 540.20$\pm0.96$ & 540.24$\pm0.97$ & 540.20$\pm0.97$ & 1.0024 \\
    $M_{\mbox{p}} (M_{\oplus})$  & 3.176$\pm0.063$ & 3.176$\pm0.063$ & 3.190$\pm0.063$ & 3.176$\pm0.060$ & 3.177$\pm0.063$ & 3.175$\pm0.062$ & 3.178$\pm0.062$ & 3.177$\pm0.062$ & 1.0003\\
    $e$ & 0.246$\pm0.028$ & 0.247$\pm0.031$ & 0.249$\pm0.029$ & 0.243$\pm0.028$ & 0.249$\pm0.030$ & 0.242$\pm0.031$ & 0.246$\pm0.031$ & 0.246$\pm0.032$ & 1.0031\\
    $ cosi $ & 0.625$\pm0.019$ & 0.623$\pm0.018$ & 0.625$\pm0.018$ & 0.626$\pm0.017$ & 0.625$\pm0.018$ & 0.624$\pm0.018$ & 0.624$\pm0.017$ & 0.625$\pm0.018$ & 1.0012 \\ 
    $M_0$ ($^{\circ}$) & 110.06$\pm6.50$ & 108.20$\pm6.26$ & 111.23$\pm7.48$ & 111.99$\pm6.28$ & 110.34$\pm7.96$ & 110.06$\pm6.77$ & 109.76$\pm7.17$ & 110.85$\pm6.95$ & 1.0130 \\
    $\epsilon_{x} (\mu\mbox{as})$ & 0.353$\pm0.060$ & 0.350$\pm0.062$ & 0.354$\pm0.061$ & 0.352$\pm0.063$ & 0.352$\pm0.065$ & 0.352$\pm0.061$ & 0.351$\pm0.063$ & 0.352$\pm0.062$ & 1.0002 \\
    $\omega$ ($^{\circ}$) & 32.54$\pm6.60$ & 210.74$\pm6.85$ & 112.68$\pm89.15$ & 156.17$\pm84.36$ & 122.21$\pm91.25$ & 152.23$\pm86.88$ & 112.58$\pm90.31$ & 70.14$\pm74.19$ & 1.2379\\
    $\Omega$ ($^{\circ}$) & 207.06$\pm2.09$ & 27.13$\pm2.99$ & 128.09$\pm89.17$ & 85.34$\pm84.41$ & 117.70$\pm90.24$ & 87.31$\pm85.06$ & 126.79$\pm89.40$ & 170.23$\pm72.68$ & 1.2455 \\
    $\omega+\Omega$ ($^{\circ}$) & 239.60$\pm6.40$ & 237.87$\pm7.39$ & 240.77$\pm7.23$ & 241.51$\pm7.58$ & 239.91$\pm7.82$ & 239.53$\pm6.61$ & 239.37$\pm8.45$ & 240.37$\pm6.81$ & 1.0109 \\
\hline
\end{tabular}
\label{tab:fit1pRc}
\end{table*}
\normalsize

\begin{table*}
\caption[Model parameters and the fitted quantities]
{Model parameters and the fitted means and one standard deviations obtained from all 8 reference stars in the single-planet model.}
\begin{tabular}{cccccccc}
\hline
\hline
        CASE & $P ({\mathrm {days}})$ & $M_{\mbox{p}} (M_{\oplus})$ & $e$ & $ cosi $ & $M_0$ ($^{\circ}$) & $\omega+\Omega$ ($^{\circ}$) & $\epsilon_{x} (\mu\mbox{as})$ \\
\hline
    Model & 541.37 & 3.100 & 0.200 & 0.643  & 100  & 230 &  \\ 
    Ref 1 & 540.22$\pm0.94$ & 3.176$\pm0.063$ & 0.246$\pm0.028$ & 0.625$\pm0.019$ & 110.1$\pm6.5$ & $239.6\pm6.4$ & 0.35$\pm0.06$ \\
    Ref 2 & 542.75$\pm0.98$ & 3.172$\pm0.061$ & 0.250$\pm0.028$ & 0.638$\pm0.018$ & 110.8$\pm6.9$ & $240.5\pm6.9$ & 0.34$\pm0.06$ \\
    Ref 3 & 539.97$\pm0.91$ & 3.249$\pm0.057$ & 0.220$\pm0.028$ & 0.604$\pm0.016$ & 102.3$\pm7.3$ & $231.4\pm7.2$ & 0.23$\pm0.08$ \\
    Ref 4 & 542.63$\pm1.04$ & 3.108$\pm0.058$ & 0.213$\pm0.032$ & 0.614$\pm0.019$ & 108.3$\pm10.4$ & $222.9\pm10.4$ & 0.39$\pm0.06$ \\
    Ref 5 & 540.31$\pm0.98$ & 3.191$\pm0.059$ & 0.203$\pm0.029$ & 0.632$\pm0.018$ & 104.0$\pm9.0$ & $233.4\pm8.8$ & 0.35$\pm0.06$ \\
    Ref 6 & 540.81$\pm0.93$ & 3.162$\pm0.061$ & 0.186$\pm0.028$ & 0.620$\pm0.017$ & 92.9$\pm8.2$ & $220.8\pm8.3$ &  0.34$\pm0.06$ \\
    Ref 7 & 540.65$\pm1.05$ & 3.061$\pm0.061$ & 0.242$\pm0.031$ & 0.658$\pm0.019$ & 106.9$\pm8.3$ & $235.8\pm8.0$ & 0.42$\pm0.06$ \\
    Ref 8 & 542.96$\pm1.02$ & 3.089$\pm0.061$ & 0.241$\pm0.033$ & 0.612$\pm0.019$ & 88.4$\pm8.3$ & $219.2\pm8.5$ & 0.47$\pm0.06$ \\
\hline
\end{tabular}
\label{tab:fit1p}
\end{table*}
\normalsize

\subsection{Single-planet system}
\label{sec:case1}

The simplest case is to infer the orbital parameters of a system with only one planet.
We simulate the differential astrometric signals of a planet with a mass of 3.1$M_{\oplus}$ and an orbital period of 541 days.
The observation strategy of our mission is to make 300 measurements of the offset angles between the target star and its reference stars over a period of 5 years.
We randomly remove several consecutive clusters from the entire time series to better simulate  real observations, which generally contain discontinuities because the targets are not always visible from the satellite.
As a consequence, we generate 8 pairs of $\sim$ 250-point time series of the angular wobbles in right ascension and declination using 8 relatively stable reference stars.
The standard deviation of the simulated Gaussian measurement error is 1 $\mu\mbox{as}$.

Using the Nii code that implements the APT-MCMC sampling strategy described in Section \ref{sec:fit}, we fit the mass and orbital parameters of the simulated single-planet system using the differential angular measurements of the 8 reference stars.
Our MCMC chains contain 1,000,000 iterations and throw the first 300,000 iterations as burn-in.
To ensure the convergence of the sampling parameters, we select one reference star, perform independent APT-MCMC runs, and calculate the Gelman-Rubin convergence diagnostic criterion $R_c$. 
Table \ref{tab:fit1pRc} gives the posterior means and the standard deviations of 8 independent Markov chains using the same observation data of one reference star and the $R_c$ values calculated from this set of independent APT-MCMC samplings.
The results indicate that the convergence is very good for $P$, $M_{\mathrm p}$, $e$, $cosi$, $M_0$, and $\epsilon_0$, and that the $R_c$ values for these 6 parameters are all $\lesssim$ 1.01.
In addition, $\omega$ and $\Omega$ do not converge separately, as they both have an $R_c$ that is greater than 1.2.
However, the summation of $\omega$ and $\Omega$ converges with $R_c$ $\approx$ 1.01.
The fitted means, standard deviations and $R_c$ values of the summation of $\omega$ and $\Omega$ in these 8 independent runs are also given in Table \ref{tab:fit1pRc}.
By comparing the theoretical differential astrometry curves corresponding to a variable $\omega$, a variable $\Omega$, and different combinations of $\omega$ and $\Omega$, we find that similar observation curves can be obtained as long as the sum of the two variables remains the same.
The argument of periapsis $\omega$ is related to the orbital eccentricity, and thus, its value has no meaning in the case of circular orbits. 
Likewise, the longitude of the ascending node $\Omega$ is related to the orbital inclination, and thus, its value is also meaningless for circular orbits.
For the magnitudes of the orbital eccentricity and inclination in our simulated model, the differences between the differential astrometry curves corresponding to a wide variety of combinations of $\omega$ and $\Omega$ are minuscule when the summation of these two variables remains unchanged. 
Therefore, in our APT-MCMC sampling process, $\omega$ and $\Omega$ yield different combinations with similar sums, thereby causing these two parameters to fail to converge individually.

Figure \ref{fig:1b} shows the corner plots \citep{Foreman-Mackey2016} of all the model parameters, among which $\omega$ and $\Omega$ are displayed in the form of their summation, obtained using the differential astrometric signals of one of the reference stars in the single-planet model.
We can see from the marginal posterior distributions that all the orbital parameters converge well.
The corner plots also show that $M_0$ is strongly correlated with the sum of $\omega$ and $\Omega$.
$M_{\mbox{p}}$ and $cosi$ are also correlated since both of these parameters determine the magnitude of the disturbance in differential astrometric signals.

Figure \ref{fig:1a} compares the simulated differential astrometric signals corresponding to the reference star used in the fitting of Figure \ref{fig:1b} and the theoretical signals generated by the marginal-posterior mean values of $M_{\mbox{p}}$, $a$, $e$, $cosi$, $P$, and $M_0$. 
The values of $\omega$ and $\Omega$ used to generate the theoretical signals are set to the values in the final iteration of the MCMC chain with $\beta = 1$.
These theoretical signals can accurately reproduce the angular wobbles in the simulated observations.
Figure \ref{fig:1a} further shows the residual plots of the fitted right ascension and declination curves.
The mean values of the residuals in the right ascension and declination directions are -0.0671 and 0.0035 $\mu\mbox{as}$, respectively, and the corresponding standard deviations are 0.9795 and 0.9536 $\mu$as.
These values are is in good agreement with the injected Gaussian observation error with a mean value of 0 $\mu$as and a standard deviation of 1 $\mu$as.

The retrieval processes using the simulated signals of the other 7 reference stars are also successful, and all the fitted results are summarized in Table \ref{tab:fit1p}.
Among all the parameters, the orbital period $P$ achieves the best-fit: 
the largest error among the 8 cases is $<$ 1.5 days, and all cases have a one standard deviation of $\sim$ 1 day.
The planetary mass $M_{\mbox{p}}$ is also accurately estimated with a maximum relative error of $\sim$ 4.8\%.
The other five orbital parameters are also reasonably derived.
Because none of the 8 simulated reference stars is accompanied by an additional astrometric disturbance, the fitted unknown measurement errors $\epsilon_{x}$ in all of the 8 cases are less than 0.5 $\mu\mbox{as}$,
which is less than the standard deviation of the injected Gaussian measurement error of 1 $\mu\mbox{as}$.

\subsection{Dual-planet system}
\label{sec:case2}

\begin{table*}
\caption[Model parameters and derived quantities]
{Model parameters and the fitted means and one standard deviations obtained from the two rounds of fitting using all 8 reference stars in the dual-planet model.}
\begin{tabular}{cccccccc}
\hline
\hline
        CASE & $P ({\mathrm {days}})$ & $M_{\mbox{p}} (M_{\oplus})$ & $e$ & $ cosi $ & $M_0$ ($^{\circ}$) & $\omega+\Omega$ ($^{\circ}$) & $\epsilon_{x} (\mu\mbox{as})$ \\
\hline
    Model & 882.04 & 25.000 & 0.100 & -0.259  & 200  & 120 &  \\ 
    Ref 1 & 882.78$\pm1.22$ & 24.61$\pm0.11$ & 0.148$\pm0.009$ & -0.269$\pm0.005$ & 199.1$\pm4.1$ & $116.9\pm4.2$ & 2.42$\pm0.09$ \\
    Ref 2 & 882.57$\pm1.24$ & 24.69$\pm0.12$ & 0.156$\pm0.009$ & -0.267$\pm0.005$ & 203.5$\pm3.7$ & $121.2\pm3.8$ & 2.45$\pm0.09$ \\
    Ref 3 & 882.43$\pm1.24$ & 24.59$\pm0.11$ & 0.144$\pm0.009$ & -0.272$\pm0.006$ & 200.8$\pm4.2$ & $118.4\pm4.2$ & 2.44$\pm0.09$ \\
    Ref 4 & 881.63$\pm1.24$ & 24.54$\pm0.12$ & 0.139$\pm0.009$ & -0.270$\pm0.005$ & 197.5$\pm4.4$ & $114.6\pm4.5$ & 2.46$\pm0.09$ \\
    Ref 5 & 882.53$\pm1.24$ & 24.61$\pm0.12$ & 0.150$\pm0.009$ & -0.268$\pm0.005$ & 200.4$\pm3.7$ & $118.1\pm3.8$ & 2.45$\pm0.09$ \\
    Ref 6 & 882.33$\pm1.25$ & 24.61$\pm0.12$ & 0.146$\pm0.009$ & -0.269$\pm0.006$ & 200.1$\pm4.2$ & $117.9\pm4.3$ &  2.44$\pm0.09$ \\
    Ref 7 & 882.01$\pm1.27$ & 24.61$\pm0.12$ & 0.146$\pm0.010$ & -0.268$\pm0.006$ & 199.9$\pm4.6$ & $117.5\pm4.7$ & 2.53$\pm0.09$ \\
    Ref 8 & 882.38$\pm1.26$ & 24.54$\pm0.12$ & 0.142$\pm0.009$ & -0.271$\pm0.006$ & 198.2$\pm4.0$ & $115.9\pm4.1$ & 2.52$\pm0.09$ \\
\hline
    Model & 541.37 & 3.10 & 0.200 & 0.643  & 100  & 230 &  \\ 
    Ref 1 & 543.45$\pm1.97$ & 3.02$\pm0.12$ & 0.361$\pm0.059$ & 0.356$\pm0.026$ & 121.0$\pm10.9$ & $256.4\pm10.2$ & 1.11$\pm0.06$ \\
    Ref 2 & 542.56$\pm1.80$ & 3.04$\pm0.11$ & 0.391$\pm0.061$ & 0.320$\pm0.027$ & 99.2$\pm8.5$ & $232.5\pm8.7$ & 1.11$\pm0.06$ \\
    Ref 3 & 540.97$\pm2.00$ & 2.96$\pm0.11$ & 0.334$\pm0.057$ & 0.325$\pm0.026$ & 110.8$\pm11.7$ & $241.2\pm11.4$ & 1.16$\pm0.06$ \\
    Ref 4 & 540.23$\pm2.00$ & 3.07$\pm0.11$ & 0.296$\pm0.050$ & 0.314$\pm0.024$ & 128.2$\pm12.0$ & $257.2\pm11.3$ & 1.14$\pm0.06$ \\
    Ref 5 & 545.91$\pm1.94$ & 3.01$\pm0.11$ & 0.346$\pm0.056$ & 0.334$\pm0.027$ & 115.1$\pm9.9$ & $252.1\pm9.8$ & 1.17$\pm0.06$ \\
    Ref 6 & 543.71$\pm2.07$ & 3.01$\pm0.14$ & 0.424$\pm0.057$ & 0.330$\pm0.029$ & 114.9$\pm8.6$ & $252.0\pm8.4$ &  1.28$\pm0.06$ \\
    Ref 7 & 543.43$\pm2.10$ & 3.13$\pm0.13$ & 0.413$\pm0.061$ & 0.304$\pm0.027$ & 107.8$\pm9.7$ & $242.0\pm9.6$ & 1.28$\pm0.06$ \\
    Ref 8 & 542.95$\pm1.80$ & 3.10$\pm0.11$ & 0.390$\pm0.052$ & 0.355$\pm0.027$ & 110.4$\pm8.6$ & $243.6\pm8.3$ & 1.13$\pm0.06$ \\
\hline
\end{tabular}
\label{tab:fit2p}
\end{table*}
\normalsize

\begin{figure*}
	\includegraphics[width=7.5in]{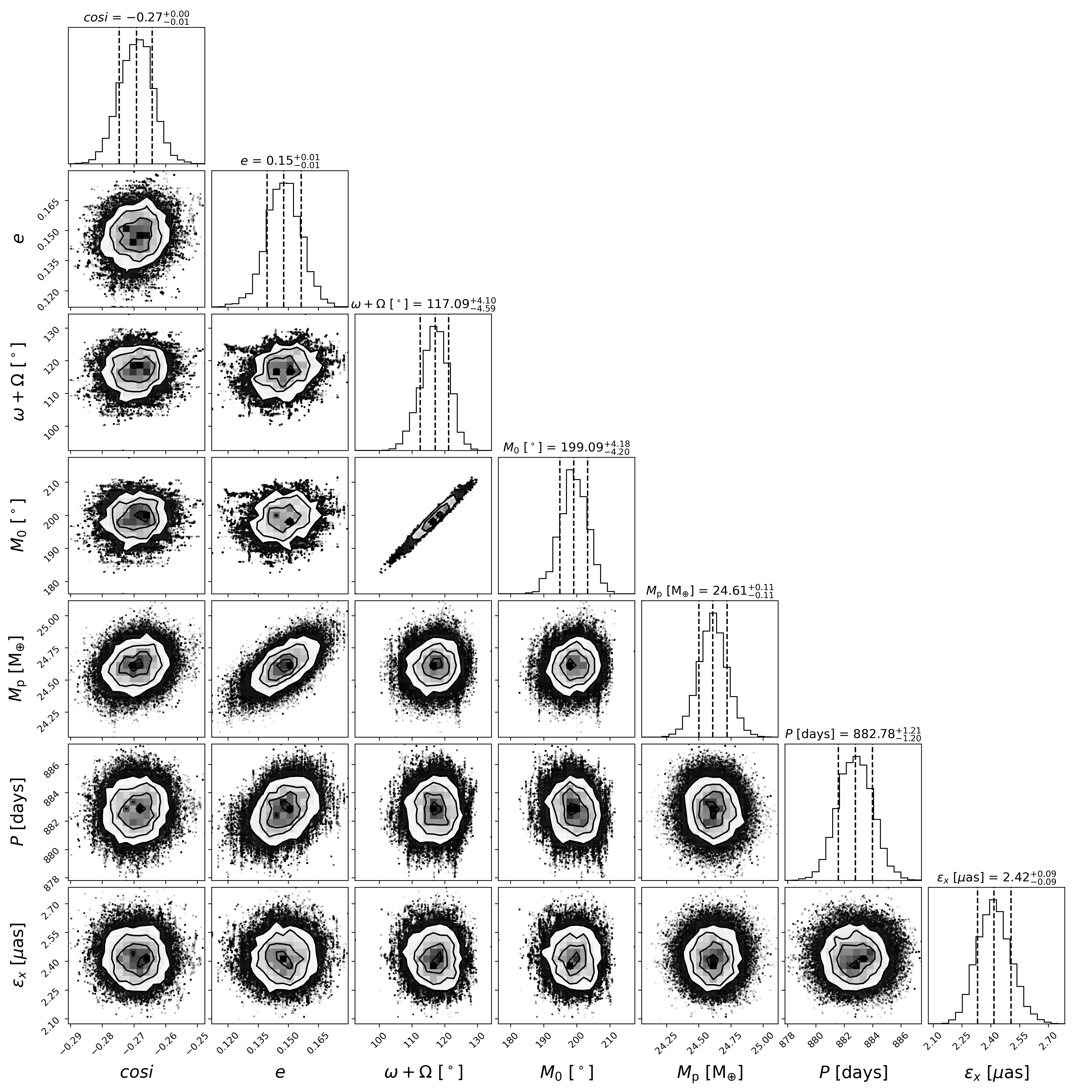}
    \caption{
        Corner plot of the posterior distributions of $a$, $e$, $cosi$, $\omega$+$\Omega$, $M_0$, $M_{\mbox{p}}$, and $P$ in the first round of fitting obtained using the differential astrometric signals of one reference star in the dual-planet model. The corresponding marginal posterior means and one standard deviation values (and these values obtained using the other 7 reference stars) are given in Table \ref{tab:fit2p}. 
    }
    \label{fig:3a}
\end{figure*}

\begin{figure*}
	\includegraphics[width=7.5in]{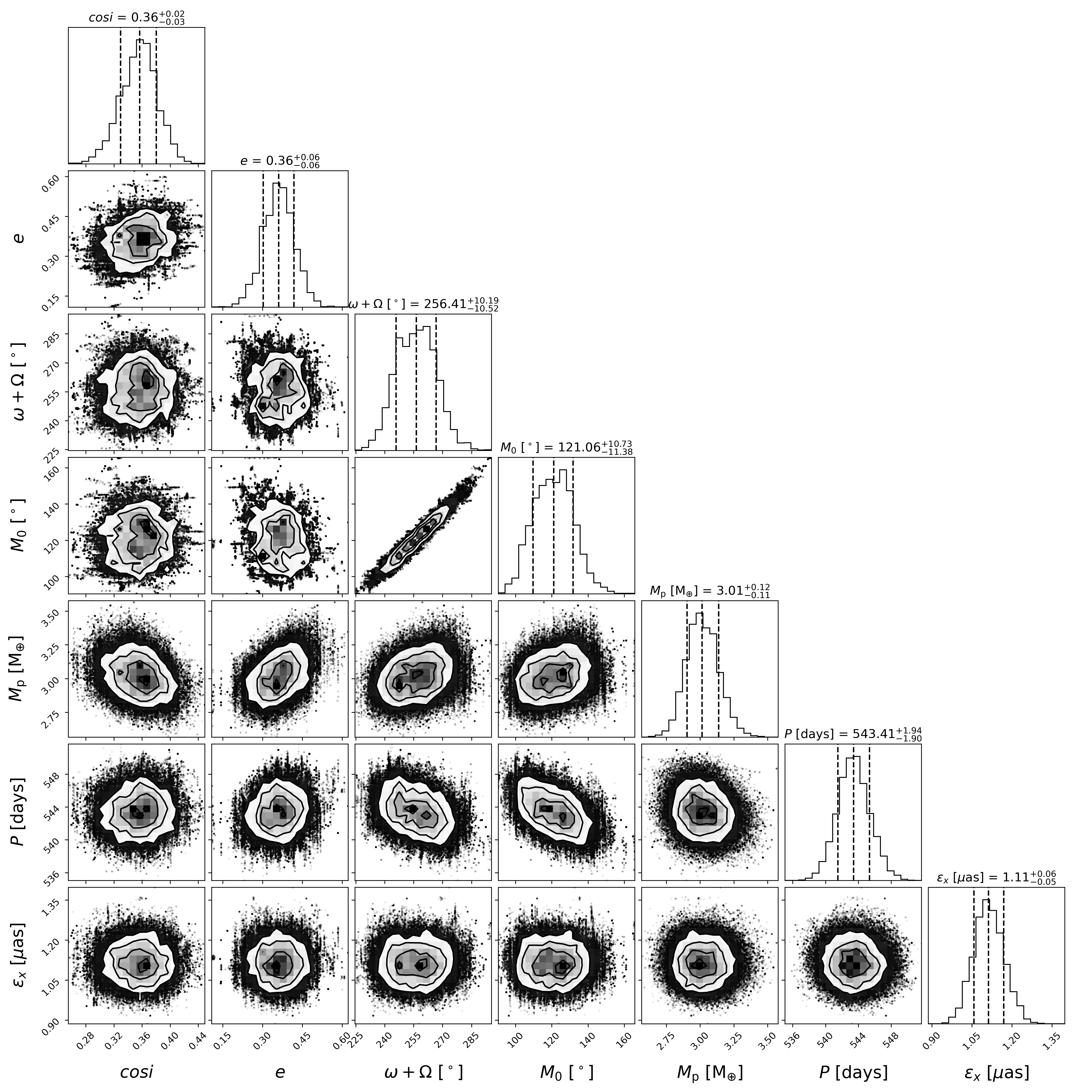}
    \caption{
        Corner plot of the posterior distributions of $a$, $e$, $cosi$, $\omega$+$\Omega$, $M_0$, $M_{\mbox{p}}$, and $P$ in the second round of fitting obtained using the differential astrometric signals of one reference star in the dual-planet model. The corresponding marginal posterior means and one standard deviation values (and these values obtained using the other 7 reference stars) are given in Table \ref{tab:fit2p}. 
    }
    \label{fig:3b}
\end{figure*}

\begin{figure*}
	\includegraphics[width=6.5in]{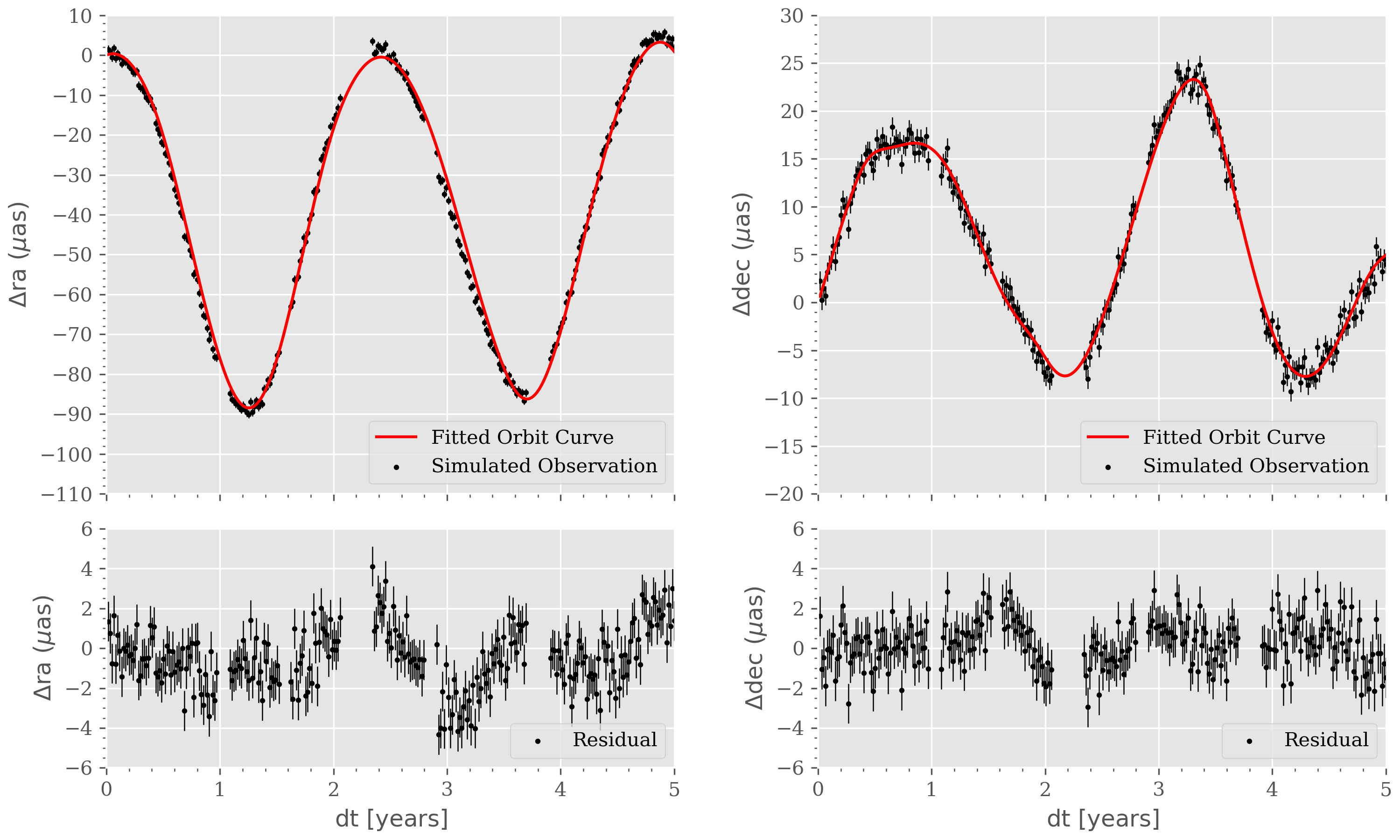}
    \caption{
        Comparison between the simulated observation of the dual-planet model and the theoretical signals generated by the fitted parameters corresponding to Figure \ref{fig:3a} and \ref{fig:3b}.
	    The points with error bars show the simulated differential astrometric observations generated using the MERCURY6 N-body integration package, and the red lines signify the theoretical signals corresponding to our fitted parameters.
            The bottom panels show the residuals. 
The residuals shown in the bottom panels contain the mutual gravitational interaction between the two planets that our fitting approach cannot explore.
    }
    \label{fig:4}
\end{figure*}

Retrieving the orbital parameters of a multiplanet system is far more complicated than retrieving those of a single-planet system, as the astrometric signals of the planets constructively or destructively interfere with one another due to their mutual gravitational interactions.
Moreover, as shown in Equation \ref{eq:alpha}, the strength of an astrometric signal is proportional to the product of a planet's mass and its orbital semimajor axis.
Therefore, when there are massive planets at distant orbits, the disturbances caused by low-mass planets at closer orbits become difficult to detect.

Here, we simulate a dual-planet system with two planets featuring different masses.
The mass and orbital parameters of the low-mass planet, as well as the characteristics of the target star, are the same as those in the single-planet system case described in Section \ref{sec:case1}.
Further, we add a massive planet of 25$M_{\oplus}$ at a larger distance into the dual-planet model;
the orbital parameters of this high-mass planet are given in Table \ref{tab:fit2p}.
In this system, the strength of the astrometric signal generated by the low-mass planet is approximately 10\% of that of the massive planet.
To include the mutual gravitational interaction between the two planets, 
the differential astrometric signal of the dual-planet system is simulated with the N-body package MERCURY6 \citep{Chambers1999}.

One correct approach to conduct the orbit retrieval of a dual-planet system is to simultaneously fit both planets' orbital elements at the same time.
Such an approach requires the combination of our APT-MCMC code with an N-body integration package.
In the Nii code, instead of a more direct approach of using an N-body integration package to simulate the dual-planet system, we implement a simplified version of this method:
we simultaneously fit both planets' orbital elements using a linear superposition of the two planets' Keplerian orbits.
This process of simultaneously fitting the orbital elements of both planets involves 15 parameters, i.e., two pairs of the elements $(a, e, i, \omega, \Omega, M_0, M_{\mbox{p}})$
and one parameter $\epsilon_{x}$ to account for additional measurement errors. 
However, we find that our APT-MCMC sampling strategy cannot converge for such a 15-parameter fitting process. 
Even if we set the initial values of the parameters near their true value, only the parameters of the massive planet can converge after 1,000,000 iterations.
The main reason may be that the disturbance of the low-mass planet is too small compared to that of the massive planet. 
Therefore, it is difficult for our automatic control system to dynamically determine proper combinations of the sampling step sizes for the orbital parameters of the two planets corresponding to their different astrometric signal strengths.
A more efficient sampling strategy is needed to achieve convergence of the two-planet model. 
We will investigate this in subsequent work.

Ultimately, we adopt a two-stage fitting process to retrieve the orbital elements of the dual-planet system. 
In the first round, we fit the two planets' signals as though they originate from a single-planet system.
As a result, the fitted unknown measurement errors $\epsilon_{x}$ are significantly larger than the standard deviation of the Gaussian measurement error. 
This implies that there may be other planets within the system.
In the second round, we subtract the signals corresponding to the fitted massive planet with the parameters obtained in the first round of fitting and perform another round of fitting on the residuals.
In both rounds, our MCMC chains contain 1,000,000 iterations and throw the first 300,000 iterations as burn-in.

Table \ref{tab:fit2p} gives the results of these two rounds of fitting for all 8 reference stars.
The orbital period $P$, $M_0$, and the sum of $\omega$ and $\Omega$ are well fitted in the first round.
For the mass $M_{\mbox{p}}$, the fitted values are approximately 1.5\% less than the injected value.
The 8 fitting processes using different reference stars all yield a larger $e$ and a smaller $cosi$.
An important indicator is the fitted unknown measurement errors $\epsilon_{x}$, which is $\sim$ 2.5 $\mu\mbox{as}$ for all 8 cases. 
Since the fitted values of $\epsilon_{x}$ are much larger than the standard deviation of the simulated Gaussian measurement error of 1 $\mu\mbox{as}$, other signals are inferred to exist within the system.
In the second round of fitting, the main parameters of the low-mass planet, i.e., its planetary mass and orbital period, are well fitted, but the relative errors of the fitted values are larger than those in the case of the single-planet system in Section \ref{sec:case1}.
Compared with the results of the single-planet system in Table \ref{tab:fit1p}, the other parameters are not well fitted, especially $e$ and $cosi$. 
We thus derive a much larger $e$ and a much smaller $cosi$.
Because the fitted values of $\epsilon_{x}$ in the second round of fitting in the 8 cases are all less than 1.3 $\mu\mbox{as}$, similar to the standard deviation of the injected Gaussian measurement error, our fitting process ends after the second round.

The reason for the insufficient fits of $e$ and $cosi$ in the dual-planet system can be explained from Figures \ref{fig:3a} and \ref{fig:3b}, which present the corner plots of all the model parameters.
Once again, $\omega$ and $\Omega$ are displayed as their summation obtained in the first and second rounds of fitting using the differential astrometric signals of one of the reference stars in the dual-planet model.
As shown in Figure \ref{fig:3a}, $M_{\mbox{p}}$ and $e$ are clearly correlated.
In the first round, we seek to fit the astrometric signals of a dual-planet system using the disturbance of a single massive planet; consequently, the disturbance caused by the low-mass planet is initially explained by adjusting the eccentricity of the orbit.
As a result, we obtain a larger orbital eccentricity and a smaller planetary mass in the first round of fitting.
These inaccurate fitting values cause the residual after the first round of fitting to contain some distorted information, which further contributes to the deviations in the orbital eccentricity and orbital inclination in the second round of fitting. 
This outcome suggests that there should be a better way to fit the orbital elements of all the planets in a multiplanet system simultaneously.
Unfortunately, our Nii code fails follow this approach, i.e., to fit all 15 orbital elements of the considered dual-planet system at the same time.
Nevertheless, since our two-stage fitting approach can accurately obtain the masses and orbital periods of the two planets, which is the main information we seek, in this work, we continue to employ this imperfect method.

Figure \ref{fig:4} compares the simulated differential astrometric signals of the dual-planet system corresponding to the reference star used in the fitting of Figure \ref{fig:3a} and \ref{fig:3b} and the theoretical signals generated by the marginal posterior mean values of $M_{\mbox{p}}$, $a$, $e$, $cosi$, $M_0$, and $P$ in combination with $\omega$ and $\Omega$ in the final iteration in the MCMC chain.
The mean values of the residuals in the right ascension and declination directions are -0.5072 and 0.1995 $\mu$as, respectively, and the corresponding standard deviations are 1.5358 and 1.1545 $\mu$as.
We can see from the residuals that although the simulated observations can be accurately explained by the fitted curves overall, clear local features remain that cannot be reproduced by the fitted curves.
These local features in the residuals originate from two aspects: 
one is the mutual gravitational action between the two planets that our fitting approach cannot explain, and the other is our inaccurate fits of the orbital eccentricity and inclination of the two planets.

\section{Summary}
\label{sec:sum}

The Nii code implements an APT-MCMC framework for the sampling of multidimensional posterior distributions and provides an observation simulation platform for the differential astrometric measurement of exoplanets.
In this paper, we test the orbit retrieval model in two cases of a single-planet system and a dual-planet system. 
In the single-planet case, all the orbital elements can be accurately fitted.
In the dual-planet case, the fitting accuracies of the planetary masses and orbital periods of the two planets can reach the same level as those in the single-planet system;
however, due to the shortcomings of our two-stage fitting approach, the fitting accuracies of the other orbital elements of the dual-planet system are lower than those of the single-planet system.
We will carry out an in-depth investigation of the orbit retrieval of multiplanet systems in a forthcoming study. 

Since Nii is an open source Python-based package, it provides an easy-to-use and expandable implementation of APT-MCMC.
To simplify the application of this code in other scientific problems with different posterior distributions, we provide many control parameters in the APT part to facilitate the adjustment of the MCMC sampling strategy,
for example, the number of parallel chains, the $\beta$ values of different chains,
the number of proposed swaps between all the parallel chains, 
the average number of iterations between each proposed swap,
the dynamic range of the sampling step sizes,
and frequency of adjusting the step sizes.
These easy-to-use control parameters ensure that the MCMC sampling strategy is sufficiently adjustable to achieve rapid convergence on a specific posterior distribution.

By adapting different prior and likelihood functions, the Nii program can be applied to different Bayesian analysis problems.
Such modifications are relatively simple in Nii.
We also plan to provide a C language version of Nii in a follow-up work for further astrophysical usage.

\section*{Acknowledgments}

We thank the referee for a thorough and constructive report that significantly improved the manuscript.
This work is supported by the B-type Strategic Priority Program of the Chinese Academy of Sciences (Grant No. XDB41000000), the Strategic Priority Research Program on Space Science of the Chinese Academy of Sciences (Grant No. XDA15020800), National Natural Science Foundation of China (Grant Nos. 11973094, 12111530175, 12033010, 11873097, 11633009), Youth Innovation Promotion Association CAS (2020319), and Foundation of Minor Planets of Purple Mountain Observatory.


\section*{DATA AVAILABILITY}
The Nii code and the test runs presented in this work is public available through GitHub.






\bsp	
\label{lastpage}
\end{document}